# Like a coin spinning in the air: the effect of (non-)metaphorical explanations on comprehension and attitudes towards quantum technology

Aletta Lucia Meinsma[1,2], W. Gudrun Reijnierse[3], Julia Cramer[1,2]

[1] Leiden Institute of Physics, Faculty of Science, Leiden University, Leiden, The Netherlands
[2] Department of Science Communication & Society, Faculty of Science, Leiden University, Leiden, The Netherlands
[3] Department of Language, Literature and Communication, Vrije Universiteit Amsterdam, Amsterdam, The Netherlands

**Abstract:** The complexity of the science underlying quantum technology may pose a barrier to its democratization. This study investigated whether metaphors improve comprehension of, and shape attitudes toward, quantum technology. In an online experiment ($n$ = 1,167 participants representative of the Dutch population), participants read a news article that included a metaphorical, non-metaphorical, or no explanation of a quantum phenomenon. Both explanation types reduced perceived comprehension of the news article compared to the control group, but increased actual comprehension of the quantum phenomenon. No direct effects were found on affect-based or cognition-based attitudes. Mediation analyses revealed a very small negative indirect effect of explanations on attitudes, through lower perceived comprehension, and a very small positive indirect effect of explanations on attitudes via increased actual comprehension – though the latter was counteracted by a negative direct effect. As metaphors offered no additional benefit over non-metaphorical explanations, the findings suggest they do not provide a communicative advantage for enhancing understanding or shaping attitudes in this context.

## 1 Introduction

Emergent technologies – defined as quickly growing, scientifically-based innovations whose market potential has not yet fully been exploited (Cozzens et al., 2010) - may feel intangible to the general public. Moreover, public awareness of such technologies is typically limited (Cobb & Macoubrie, 2004; Scheufele & Lewenstein, 2005). Public reactions to previous emergent technologies have shown that public resistance can arise (Kurath & Gisler, 2009). This highlights the importance to take early account of how the general public perceives a new technological development (Mooney, 2010). Engaging the public early in a technology's development can help scientists better understand people's concerns, before a conflict arises (Mooney, 2010).

An example of a current emergent technology is quantum technology. Quantum technology encompasses three domains: quantum computing and simulation, quantum communication, and quantum sensing and metrology (Stichting Quantum Delta NL, 2020). This technology holds the promise of significant societal benefits. For instance, quantum technology promises *"a new window into the underground"* (p. 1), thereby reducing the risk of encountering unexpected ground conditions during the construction of vital energy, transportation, and utility infrastructure (Stray et al., 2022). It may also assist in improving computations that are important for the development of new drugs (Outeiral et al., 2021), and promises to enable fundamentally secure communications globally (Wehner et al., 2018). In contrast, quantum technology raises potential concerns and ethical implications, for instance through the question of who gets access and who falls behind (Ten Holter et al., 2022), and through its potential for criminal use (Vermaas et al., 2019). In addition, the science behind quantum technology is counterintuitive to what we experience in our daily lives, which makes it difficult to grasp.

One way that may make the science behind quantum technology easier to understand is through the use of metaphors. A well-known example is Schrödinger's cat[1] (Van de Merbel et al., 2024), which uses the metaphor of a cat in a box - where the cat is both dead and alive at the same time until the box is opened - to describe the quantum phenomenon of particles existing in two states simultaneously. By comparing an abstract quantum phenomenon (the target domain) to a cat in a box (a more concrete source domain), metaphors are used to bridge the gap between complex quantum phenomena and everyday experiences.

Metaphors are for instance used in newspaper articles about quantum science/technology (Hilkamo & Granqvist, 2022), which are important sources for people to get to know about quantum technology (European Commission, Directorate-General for Communication, 2021; Van de Merbel et al., 2024). However, to our knowledge, little is known about the effects of such metaphors on news media recipients.

In this article, we examined the effects of (non-)metaphorical explanations of counterintuitive quantum phenomena in a news article about quantum technology. Specifically, we studied whether (non-)metaphorical explanations influence people's comprehension and attitudes towards the technology.

## 2 Theory

As science and technology are often abstract, complex and unfamiliar, metaphors may make them more concrete, simpler, and easier to understand (Lakoff & Johnson, 1980). According to Lakoff & Johnson (1980) metaphors are a mapping between different domains (note that based on this definition, analogies and similes also fall under the term 'metaphor', which we have therefore also referred to as metaphor throughout this article). Metaphors in communication of science and technology towards a general public generally function to explain a scientific concept in terms of a more familiar one (Beger & Smith, 2020; Smedinga et al., 2023). In this way, metaphors could be helpful in making the complex field of quantum technology more accessible.

*2.1 Incomprehensible quantum science could limit public engagement*

There is a call for making quantum technology more accessible and comprehensible to a general public (Coenen et al., 2022) and engage them early on in its development process (Roberson et al., 2021). One of the reasons for this is that involving a wider group of people in envisioning the impact of quantum technology fits a more democratic process, in which citizens can express their opinions and concerns about developments that may affect them (Van Dam et al., 2020). Moreover, engaging the public in quantum technology may contribute to more public support and less resistance towards it, with previous emergent technologies such as nuclear energy and biotechnology having to deal with public resistance (Druckman & Bolsen, 2011; Kurath & Gisler, 2009).

However, the science that underlies quantum technology is complex, which could draw a barrier for public engagement. This may be reinforced by experts who emphasize this complexity in their outreach. Using Richard Feynman's well-known remark *"I think I can safely say that nobody understands quantum mechanics"* (Feynman, 1967, p. 129), experts tend to focus on quantum mechanics as something that is incomprehensible (Seskir et al., 2023). This focus could be bad for the democratization of quantum technology, as it could hinder comprehension and engagement in discussions on its impact on society (Coenen et al., 2022; Seskir et al., 2023). Research on public engagement with quantum technology found that respondents in The Netherlands felt they had little influence over the development of quantum technology (Van de Merbel et al., 2024), and in the UK participants felt quantum technology was complicated and for experts only (Busby et al., 2017).

An open question in communication about quantum technology is what types of information allow for a good connection between quantum and society. As metaphors may influence comprehension (section 2.2) and attitudes (section 2.3), together with the fact that comprehension may influence attitudes (section 2.4), metaphors are an important tool to study.

*2.2 Effect of metaphor on comprehension*

One of the functions of metaphors in science is to make scientific topics easier to understand (Beger & Smith, 2020). However, metaphors have been found to create differences between what someone feels they have understood from a text (perceived comprehension) and what they actually understood (actual comprehension, see Jaeger & Wiley, 2015; Wiley et al., 2018). Specifically, metaphors can cause people to believe that their understanding of a text is higher than it actually is (i.e., illusion of comprehension, Jaeger & Wiley, 2015; Wiley et al., 2018). Jaeger & Wiley (2015), for instance, found that students who read a text about the greenhouse effect were generally overconfident in their understanding, but those in the metaphor-enhanced text condition were even more overconfident than the ones in the control group.

When looking specifically at actual comprehension, metaphors may not necessarily improve it. While in the field of education some studies have reported positive effects, with metaphor-enhanced texts supporting recall (Glynn & Takahashi, 1998) and reasoning (Yanowitz, 2001), other studies found no significant effects with the metaphor-enhanced texts resulting in similar comprehension results as control texts (Alexander & Kulikowich, 1991; Braasch & and Goldman, 2010). Metaphors can even have detrimental effects on comprehension when learners map features between source and target domains that cannot be mapped, leading to misconceptions (Zook & Maier, 1994).

While metaphors in scientific educational texts have been well-studied, it remains unclear whether - and how - metaphors influence comprehension in science communication. A study in environmental communication by Reijnierse et al. (work in progress) showed that sustainability metaphors increased people's perceived comprehension of the text: people in the metaphor condition rated their comprehension significantly higher than those in the control group. However, there was no significant difference between the groups in terms of actual comprehension.

### 2.3 Effect of metaphor on attitude

Meta-analyses by Sopory & Dillard (2002), Van Stee (2018) and O'Keefe & Hoeken (2021) found that metaphors have a small positive effect on attitudes compared to literal messages. Metaphors with a familiar target domain furthermore led to more positive attitudes than those with an unfamiliar target domain, likely because the latter requires too much cognitive effort to process (Sopory & Dillard, 2002; Van Stee, 2018). The effect of metaphors on attitudes, however, appear to be variable, as a next, individual study may also find a negative effect size instead of a positive one (O'Keefe & Hoeken, 2021).

While these meta-analyses included studies from contexts such as advertising, politics, environmental issues and health, none of them included studies focusing on new technology. To our knowledge, the effect of metaphor on attitudes towards new technology remains unclear. Prior research has shown that even when information is very limited, people can form attitudes towards new technology instantly (Druckman & Bolsen, 2011; Van Giesen et al., 2015). These attitudes may be based more on affect – i.e., emotions and feelings towards the technology, or cognition – i.e., thoughts and beliefs towards the technology. For unfamiliar technologies, people tend to rely more on affect, whereas this changes over time when cognition starts to play a bigger role (van Giesen et al., 2018). Examining both affective and cognitive responses to an emergent technology can therefore form a good image of people's attitudes.

### 2.4 Effect of comprehension on attitude

Research suggests that comprehension is important in attitude formation and change (Wyer Jr. & Shrum, 2015). For example, comprehension of information that creates a visual image in one's mind, as well as information that is presented in the form of a story, can have a stronger effect on one's attitude. Even when the features that make a text more vivid and therefore easier to remember are completely irrelevant to the attitude in question - as in determining one's guilt in a court case after a delay in viewing the evidence - they can still influence attitude formation (Wyer Jr. & Shrum, 2015).

For emergent technology specifically, people's beliefs of having enough information to form a judgment may influence their attitudes more than what they actually know (Akin et al., 2021). In an online survey in the US, Akin et al. (2021) found that people's perceptions of their own knowledge significantly predicted their positive attitudes towards three emergent technologies (nuclear energy, nanotechnology and synthetic biology). In contrast, people's actual knowledge only predicted two of those technologies – for the newest of the three technologies, synthetic biology, actual knowledge did not significantly predict attitudes. This provides reasons to investigate the relationship between comprehension and attitudes in the context of emergent quantum technology.

*2.5 Metaphors in communication about quantum technology*

Metaphor use has already been specifically recommended in public communication about quantum technology (Grinbaum, 2017; Hilkamo & Granqvist, 2022). Grinbaum (2017), for instance, has emphasized from a philosophical viewpoint the importance of effectively explaining counterintuitive quantum phenomena to ensure the public can grasp the basics of what quantum physicists work with daily. Without this comprehension, people's perception of quantum technology may become more negative. To address this, Grinbaum (2017) suggests metaphors to explain the science and at the same time convey the beauty of quantum science. Hilkamo & Granqvist (2022) have also emphasized the importance of metaphors in the quantum technology domain, as they can appeal to emotions.

It is not yet known exactly how metaphors influence comprehension and attitudes in public communication such as news articles about quantum technology. Our research therefore aims to explore whether a (non-)metaphorical explanation of a quantum phenomenon in a news article affect comprehension (both perceived and actual) and shape attitudes (both cognition- and affect-based).

Prior studies worked with metaphor-enhanced texts, where the metaphor was added to the original explanation (see e.g., Alexander & Kulikowich, 1991; Braasch & and Goldman, 2010; Glynn & Takahashi, 1998; Jaeger & Wiley, 2015; Yanowitz, 2001). However, this results in a double explanation of the phenomenon – once non-metaphorically and once metaphorically. Because we are interested in whether a metaphorical explanation in itself provides benefits, we work with three different conditions. Our research questions are as follows (see Figure 1 below):

> RQ1: To what extent does the use of a non-metaphorical quantum phenomenon explanation vs a metaphorical quantum phenomenon explanation vs a control in a news article influence:

>> a) perceived comprehension[2]

>> b) actual comprehension of a quantum phenomenon

>> c) affect-based attitudes towards quantum technology

>> d) cognition-based attitudes towards quantum technology

> RQ2: To what extent does perceived comprehension mediate the possible effects of explanation type (non-metaphorical quantum explanation, metaphorical quantum explanation, no explanation) on affect- and cognition-based attitudes?

> RQ3: To what extent does actual comprehension mediate the possible effects of explanation type (non-metaphorical quantum explanation, metaphorical quantum explanation, no explanation) on affect- and cognition-based attitudes?

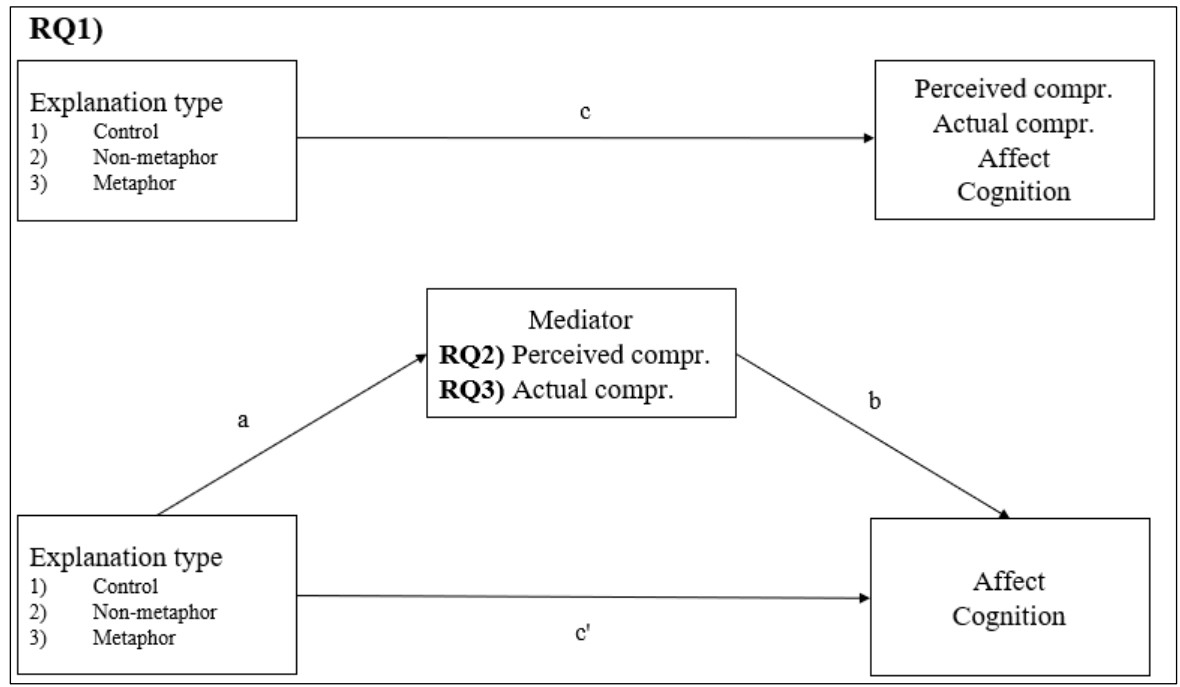

**Figure 1.** Visual representation of the research questions. The top path diagram illustrates RQ1, examining the direct relationships between the variables. The bottom path diagram shows the two mediation models: RQ2 examines perceived comprehension as a mediator, and RQ3 examines actual comprehension as a mediator.

# 3 Methodology

Following our research questions, we performed an online experiment. The study information, design plan, sampling plan, variables and analysis plan were pre-registered on the Open Science Framework before the data collection started (https://tinyurl.com/mr3hp4w9). The Ethics Review Committee of the Faculty of Science, Leiden University gave ethical approval to conduct the study (reference number 2023-021). The following section gives an overview of the materials used, the design and procedure, the participants and the variables of the study.

*3.1 Materials*

We designed a newspaper article about a new quantum computer that the European Commission has decided to build in the Netherlands as stimulus material. The article was based on an actual event (see NOS, 2024). The text contained 162 words in the control conditions, with an additional paragraph of 73 to 80 words included in the explanation conditions. The text of the news article and different conditions can be found in Appendix A.1.

We chose to explain the quantum phenomena *superposition* or *entanglement* in the explanation conditions. These phenomena are relevant to study as they underlie quantum technology, are counterintuitive from the perspective of everyday experiences and are often explained to general audiences in popular communication (Meinsma et al., 2023, 2025). To identify which metaphor for superposition and entanglement were deemed most accurate for communicating with non-experts, we conducted a small study among Dutch-speaking quantum experts.

### 3.1.1 Expert insights: method

We emailed an online survey to 67 researchers from Dutch universities, who were selected based on their university profiles. Only those working in a field where quantum physics plays a role and with Dutch as a native language could participate. First, we asked participants to provide consent. Then, after indicating that Dutch was (one of) their native language(s) and indicating the quantum-related field in which they worked (based on Fox et al., 2020), participants were randomly assigned to read five metaphors about a quantum phenomenon (superposition or entanglement) that were displayed in a random order. The metaphors were generated by ChatGPT 3.5, as this AI tool is widely used and known for its accessibility and as science communication research encourages investigating the accuracy of AI-generated content (Schäfer, 2023). The precise wordings per metaphor can be found in Table A.3 in the Appendix.

After reading each metaphor, participants rated its accuracy and validity on a 7-point Likert scale. They also indicated how likely it would be for them to use a metaphor when talking to a non-quantum expert about superposition or entanglement (7-point scale). Next, participants ranked which metaphor they would use if they were to use a metaphor in a conversation with non-quantum experts from most likely to least likely and justified their ranking. Finally, we asked participants if they knew of an alternative metaphor about quantum superposition or entanglement. If they answered *"yes"*, they were prompted to shortly describe the alternative metaphor, and rate it on accuracy and validity.

### 3.1.2 Expert insights: results

The survey ran between January 15th and March 15th, 2024 resulting in a total of $n$ = 22 completed surveys. The majority of the participants were scientists working in a field involving quantum physics ($n$ = 21), and one participant is an engineer. In total, 12 participants evaluated five metaphors about superposition, and 10 participants evaluated five metaphors about entanglement. The median completion time was 14 minutes 50 seconds. Experts indicated it was very likely ($M$ = 5.68, $SD$ = 1.36) that they would use a metaphor in a conversation with non-quantum experts to explain superposition or entanglement.

Table 1 shows that the coin metaphor scored highest for superposition ($M$ = 4.13, $SD$ = 1.86) and the dice metaphor for entanglement ($M$ = 5.30, $SD$ = 1.89). A complete overview of the experts' feedback per metaphor can be found in Table A.4 in the Appendix. Based on these results, the metaphorical explanation condition thus included the coin metaphor for superposition and the dice metaphor for entanglement.

**Table 1.**
Mean total scores (and SDs between brackets) for superposition and entanglement answered on a 7-point Likert scale. The total scores are an average of the scores on accuracy and validity, and are ordered from highest to lowest. The precise wordings per metaphor can be found in Table A.3 in the Appendix.

| Superposition | | |
|---|---|---|
| | Metaphor | Total |
| 1 | A coin spinning in the air | 4.13 (1.86) |
| 2 | A radio producing a jumble of sounds | 4.04 (2.94) |
| 3 | A cat, a vial of poison and a radioactive atom in a locked box | 3.21 (2.59) |
| 4 | An artist dabbing his brush in multiple colours | 3.04 (2.31) |
| 5 | A musician composing a music piece | 2.25 (2.01) |
| Entanglement | | |
| | Metaphor | Total |
| 1 | A pair of dice | 5.30 (1.89) |
| 2 | Two dancers performing a perfectly synchronized dance routine | 3.80 (2.55) |
| 3 | A telepathic twin | 3.35 (2.06) |
| 4 | Two compass needles that always point in opposite directions | 3.15 (2.46) |
| 5 | Two clocks with perfectly synchronized second hands | 2.90 (2.53) |

*3.2 Design and procedure*

The main study made use of an experimental design with 6 conditions: 1 factor with 3 levels (explanation type: metaphorical, non-metaphorical, no explanation) and 2 items (quantum phenomena: superposition and entanglement). Participants provided consent and then declared to not use any external sources, as we wanted to explicitly discourage the use of search engines and AI (see Meem et al., 2024). General information (age, gender, education) and control variables were asked, after which participants were randomly assigned to one of the 6 conditions. After reading the news article, participants were asked to answer questions on their perceived comprehension, actual comprehension, affect-based and cognition-based attitudes.

*3.3 Participants*

The experiment ran between January 29th and February 3rd, 2025 and resulted in a total sample of *n* = 1,176 participants. Participants were recruited by PanelClix (https://www.panelclix.nl/), an external panel service in the Netherlands. For each completed questionnaire in PanelClix, participants receive points ('clix') which they can exchange for money amongst others. The median time to complete the survey was 5 minutes and 50 seconds.

A total of *n* = 9 participants were excluded from the full analyses because they a) finished the survey within 90 seconds (*n* = 4); b) answered both the attention check question and the actual comprehension question incorrectly (*n* = 3); or c) reported being under the age of 18 (*n* = 2). As 36 participants failed to answer the actual comprehension question by mainly typing a letter or a non-existing word, they were excluded from only that part of the analysis. This resulted in a final sample of *n* = 1,131 participants for the analyses involving the actual comprehension variable and 1,167 participants for the remaining analyses.

The sample characteristics of the 1,167 participants closely matched the population statistics of the Netherlands (CBS, 2024). In terms of gender, $n = 596$ (51.1%) identified as male, $n = 568$ (48.7%) as female and $n = 3$ (0.3%) as other. Participants were between 18 and 88 years old ($M = 47.7$, $SD = 16.8$). In terms of education level, $n = 236$ participants (20.2%) reported having completed a low level of education, $n = 591$ (50.6%) reported an average level, and $n = 340$ (29.1%) reported a high level of education.

*3.4 Dependent variables*

**Perceived comprehension.** Perceived comprehension of the newspaper article was measured with 3 items (Miele & Molden, 2010). These are: *"How well do you feel you understand the news article?"* [1 = very poorly, 7 = very well], *"How certain are you that you will answer questions correctly about the news article?"* [1 = very uncertain, 7 = very certain], and *"How confused about the news article do you feel?"* [1 = not at all confused, 7 = very confused] (reverse coded). The three items were averaged into an index ($M = 4.38$, $SD = 1.20$, Cronbach's alpha = 0.73).

**Actual comprehension.** To check whether participants remembered which quantum phenomenon was mentioned in the news article, we asked a control question. Participants were asked to select the phenomenon they had just read from a multiple-choice question (answer options: superposition, entanglement, tunnelling, decoherence, I don't know), and based on their answer, were asked to write in their own words what the selected phenomenon meant. If 'I don't know' was selected, the question was formulated as: *"Describe in your own words how particles behave at the smallest scale"*. We calculated the index on a scale from 0 to 4 based on the text provided by the participants ($M = 0.43$, $SD = 0.78$). A point was given for each of the following elements:

● For superposition: 1) mention the connection with quantum; 2) something can be A and B at the same time; 3) a measurement has influence, 4) after which a single state is left

● For entanglement: 1) mention the connection with quantum; 2) there is a connection/correlation; 3) measurement of one part influences the other part; 4) which is not dependent on the distance

In addition to assigning points, participants' reuse of the metaphor was analysed. Both the index and the metaphor reuse were coded reliably, as intercoder agreement between two coders was perfect to near-perfect for almost all cases (see section A.2 in the Appendix for full details on the intercoder reliability analysis).

**Affect-based attitude.** Affect-based attitude was measured with 7 items (van Giesen et al., 2018). These are: *"How do you feel about quantum technology after having read the news article? I feel…"* *…joy, …desire, …fascination, …satisfaction, …fear* (reverse-coded), *…sadness* (reverse-coded), *…disgust* (reverse-coded) [1 = not at all, 7 = very much]. The seven items were averaged into an index ($M = 4.34$, $SD = 0.92$, Cronbach's alpha = 0.71).

**Cognition-based attitude**. Cognition-based attitude was measured with 7 items (van Giesen et al., 2018). These are: *"What is your view on quantum technology after having read the news article? Quantum technology is…."* *…useful, …functional, …beneficial, …useless* (reverse-coded), *…harmful* (reverse-coded), *…disadvantageous* (reverse-coded), *…unusable* (reverse-coded) [1 = not at all, 7 =

very much]. The seven items were averaged into an index ($M$ = 4.67, $SD$ = 0.99, Cronbach's alpha = 0.81).

*3.5 Control variables*

We further measured five control variables, since we expected these to potentially influence our dependent variables. These were: *awareness of quantum*, *science news use*, *interest in new technology*, *faith in intuition* and *need for cognition*. All information regarding these variables is available in the Appendix A.3.

*3.6 Analysis and statistical procedures*

Data were analysed with jamovi 2.3.28. Results of the two items (superposition and entanglement) were collapsed for the analysis to make generalizable claims on the effectiveness of (non-)metaphorical explanations across quantum phenomena. Randomization checks were performed to make sure that the participants were evenly distributed across the three conditions.[3] Participants were found to be evenly distributed across the three conditions for gender ($\chi^2$(2)=1.79, $p$ = 0.408), age ($\chi^2$(6)=3.98, $p$ = 0.679), level of education ($\chi^2$(4)=2.86, $p$ = 0.581), quantum technology awareness ($F$(2, 1164) = 0.360, $p$ = 0.698), science news use ($\chi^2$(2)=0.608, $p$ = 0.738), interest in new technology ($F$(2, 1164) = 0.808, $p$ = 0.446), faith in intuition ($F$(2, 1164) = 1.04, $p$ = 0.353) and need for cognition ($F$(2, 1164) = 0.110, $p$ = 0.896). Therefore, we did not include any of these variables as a covariate in the analyses.

To answer RQ1, four separate ANOVAs were performed using the explanation condition as independent variable and perceived comprehension, actual comprehension, affect-based attitudes, and cognition-based attitudes as dependent variables.

To answer RQ2 and RQ3, four separate mediation analyses were conducted using the jAMM module. We used the explanation type condition as the independent variable, perceived comprehension (for RQ2) and actual comprehension (for RQ3) as the mediator, and affect-based attitudes and cognition-based attitudes as the dependent variable. Confidence intervals were computed with the Bootstrap percentiles method using 5000 bootstraps.

# 4 Results

*4.1 Main effects on the dependent variables*

Statistically significant differences between the conditions were found for perceived comprehension ($F$(2,1164) = 8.49, $p$ <.001, $\eta^2$= 0.014) and actual comprehension ($F$(2,1131) = 32.6, $p$ <.001, $\eta^2$= 0.055). We found no statistically significant differences between the conditions for affect-based attitudes ($F$(2,1164) = 1.48, $p$ = 0.23) or cognition-based attitudes ($F$(2,1164) = 1.81, $p$ = 0.16).

**Table 2.**
Means (and SDs between brackets) by condition, based on the estimated marginal means estimated from the statistical model. Perceived comprehension, affect and cognition were measured on a 7-point Likert scale. Actual comprehension was measured on a scale from 0 to 4.

| | | | | Comprehension |
| --- | --- | --- | --- | --- |
| Condition | Perceived | 95% CI | Actual | 95%CI |
| Control | 4.58 (0.06) | [4.46, 4.69] | 0.19 (0.04) | [0.11, 0.26] |
| Non-metaphorical explanation | 4.26 (0.06) | [4.14, 4.38] | 0.62 (0.04) | [0.54, 0.70] |
| Metaphorical explanation | 4.29 (0.06) | [4.17, 4.40] | 0.49 (0.04) | [0.42, 0.57] |
| | | | | Attitude |
| | Affect | 95% CI | Cognition | 95% CI |
| Control | 4.39 (0.05) | [4.30, 4.48] | 4.74 (0.05) | [4.64, 4.84] |
| Non-metaphorical explanation | 4.36 (0.05) | [4.26, 4.45] | 4.66 (0.05) | [4.56, 4.76] |
| Metaphorical explanation | 4.28 (0.05) | [4.19, 4.37] | 4.61 (0.05) | [4.51, 4.71] |

Results of the pairwise comparisons with a Bonferroni correction showed that for perceived comprehension, the control group scored significantly higher than both the non-metaphorical group (mean difference = 0.32, $SE$ = 0.09, $p$ <.001) and the metaphorical group (mean difference = 0.29, $SE$ = 0.08, $p$ = 0.002). There was no significant difference between the metaphorical and non-metaphorical group (mean difference = -0.028, $SE$ = 0.09, $p$ = 1.000). For actual comprehension, the control group scored significantly lower than the non-metaphorical group (mean difference = -0.43, $SE$ = 0.06, $p$ <.001) and the metaphorical group (mean difference = -0.31, $SE$ = 0.05, $p$ = <.001). The difference between the non-metaphorical group and the metaphorical group was not significant (mean difference = 0.13, $SE$ = 0.06, $p$ = 0.069).

*4.2 Mediation effects*

Although no significant differences were found between the conditions on affect-based and cognition-based attitudes, mediation can still exist in the absence of a main effect (O'Rourke & MacKinnon, 2018). In performing the mediation analyses, we chose the control group as a reference group given the outcomes of the main effect analyses.

*4.2.1 Perceived comprehension as mediator*

We found some evidence that perceived comprehension acts as a mediator between explanation type and both affect-based attitudes and cognition-based attitudes, as none of the confidence intervals for the tested indirect effects contained zero. Figure 2 shows the results (see Table A.1 in the Appendix for the complete table).

As shown in Figure 2, we found a small but statistically significant decrease in perceived comprehension when participants received the non-metaphorical (β = -0.12, $p$ <.001) or metaphorical explanations (β = -0.11, $p$ < .001), compared to the control group. This indicates that both types of explanations lowered participants' feeling of understanding the news article. We furthermore found a strong statistically significant increase in both affect (β = 0.55, $p$ < .001) and cognition (β = 0.46, $p$ < .001) when participants scored higher on perceived comprehension. This indicates that there is a relationship between a higher perceived comprehension of the news article and feeling and viewing quantum technology more positively.

Given these two findings, the indirect effects found between explanation type and affect or cognition, with perceived comprehension as the mediator, were statistically significant ($p < .001$) but very small (β's ranged between -0.07 and -0.05). The direct effects, i.e. the effect of explanation type on affect or cognition when accounting for perceived comprehension, and the total effects were non-significant in all cases. This suggests that the effect of explanations on people's affect-based and cognition-based attitudes towards quantum technology is entirely dependent on how well they feel they understood a news article on the topic. Without a change in perceived comprehension, there would be no effect.

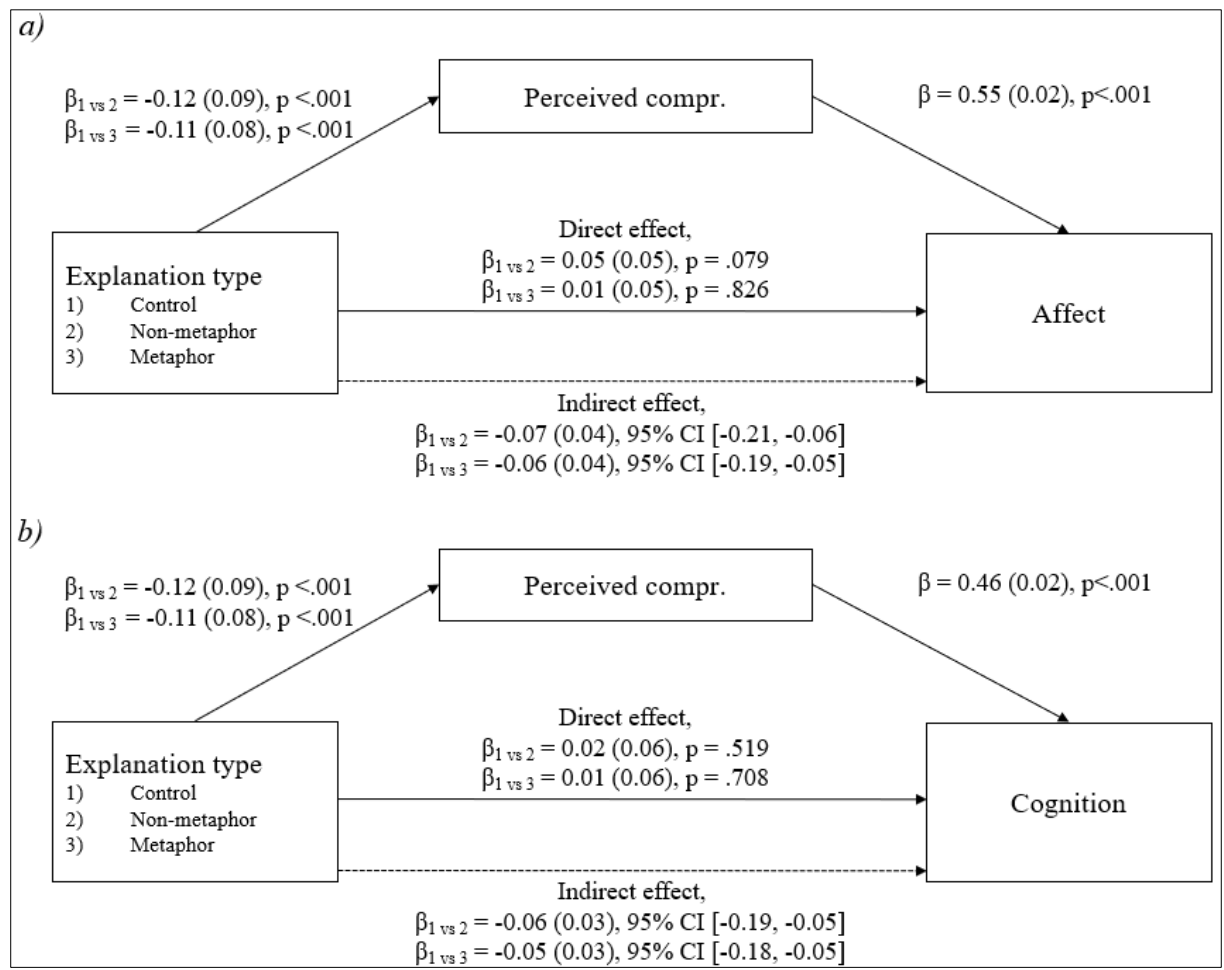

**Figure 2.** Visual representation of the results, depicting perceived comprehension as a mediator, with (a) affect and (b) cognition as dependent variables. Note that $\beta_{1 \, vs \, 2}$ ($\beta_{1 \, vs \, 3}$) denotes the effect size of the non-metaphorical group (metaphorical group) with control as reference.

### 4.2.2 Actual comprehension as mediator

We also found some evidence that actual comprehension acts as a mediator between explanation type and both affect-based attitudes and cognition-based attitudes, as again none of the confidence intervals for the tested indirect effects contained zero. Figure 3 shows the results (see Table A.2 in the Appendix for the complete table).

We found a small but statistically significant increase in actual comprehension when participants received the non-metaphorical (β = 0.26, $p <.001$) or metaphorical explanations (β = 0.19, $p < .001$), compared to the control group. We furthermore found a small statistically significant increase in both affect-based attitudes (β = 0.28, *p* < .001) and cognition-based attitudes (β = 0.27, *p* < .001) from actual comprehension. This indicates that there is a relationship between a higher actual comprehension of the quantum phenomenon and feeling and viewing quantum technology more positively.

Given these two findings, the indirect effects between explanation type and affect or cognition, with actual comprehension as the mediator, were statistically significant (*p* < .001) but very small (β's ranged between 0.05 and 0.07). The direct effect, which is the remaining effect after accounting for actual comprehension as a mediator, revealed a small negative effect (β's between -0.11 and -0.09). The total effects were non-significant. Since the indirect and direct effects have opposing signs, finding a mediating effect in this case while the total effect is non-significant is called inconsistent mediation (O'Rourke & MacKinnon, 2018). This suggests that while explanations may increase people's affect-based and cognition-based attitudes towards quantum technology by improving their understanding of quantum phenomena (i.e., actual comprehension), this positive effect is counteracted by a direct negative effect of explanation type on affect-based and cognition-based attitudes. There may be other mechanisms at play that result in such a counteracting effect.

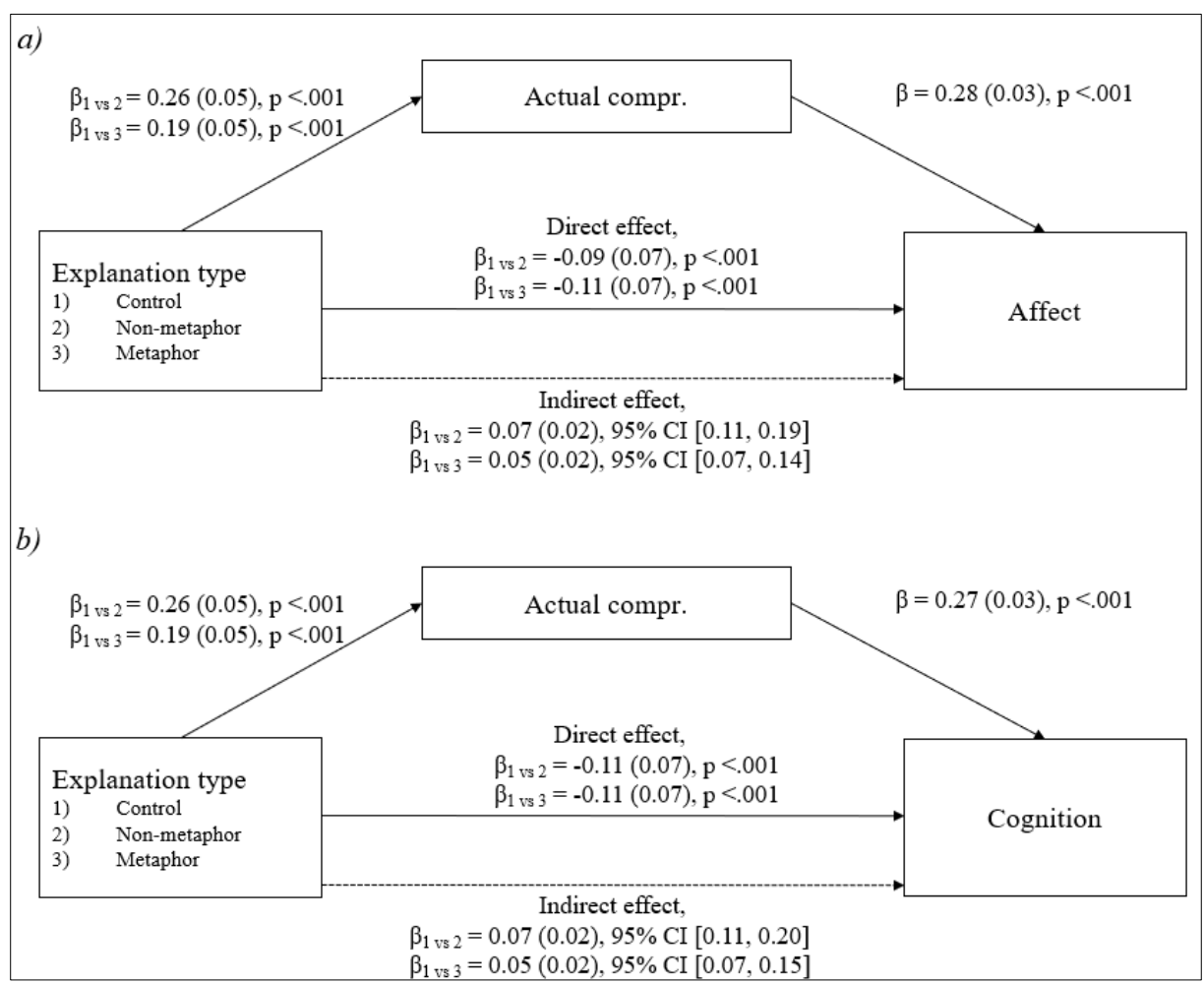

**Figure 3.** Visual representation of the results, depicting actual comprehension as a mediator, with (a) affect and (b) cognition as dependent variables. Note that $\beta_{1\,vs\,2}$ ($\beta_{1\,vs\,3}$) denotes the effect size of the non-metaphorical group (metaphorical group) with control as reference.

# 5 Discussion

This study examined, through a large-scale experiment, how (non-)metaphorical explanations of a quantum phenomenon influenced people's perceived comprehension of a news article about quantum technology, their actual comprehension of a quantum phenomenon, and their attitudes towards this emergent technology.

## 5.1 Effects of explanation type on comprehension

We found that including an explanation of a quantum phenomenon – regardless of whether it was metaphorical or non-metaphorical – led to lower perceived comprehension scores of the text, while actual comprehension scores of the quantum phenomenon were higher. The findings that metaphors can lead to higher perceived comprehension (Reijnierse et al., work in progress) or enhance people's overconfidence (Jaeger & Wiley, 2015; Wiley et al., 2018) was thus not apparent in our study. The opposite even emerged, where (non-)metaphorical explanations had a detrimental effect on participants' perceived comprehension, but improved their actual comprehension, compared to a control condition. It should be noted that actual comprehension scores were low in all conditions (≤0.62), suggesting that people did not grasp the phenomenon well.

It is important to note that we measured perceived comprehension for the text as a whole, but actual comprehension specifically for the quantum phenomenon. Perhaps the fact that explanations reduce perceived comprehension of the text indicates an information overload, where the length and amount of information make a text seem more complex. In contrast, we found that explanations contribute to an improved actual comprehension of a quantum phenomenon. This makes sense, because without an explanation it is conceivable that people do not know what superposition or entanglement entails (also note the low awareness for quantum technology, see Figure A.1), and therefore cannot answer a question about it correctly.

We found no differences between metaphorical and non-metaphorical explanations on actual comprehension, consistent with Alexander & Kulikowich (1991) and Braasch & Goldman (2010), but contrasting others (Glynn & Takahashi, 1998; Yanowitz, 2001; Jaeger & Wiley, 2015). A main difference is that we ensured the phenomenon was explained only once in the metaphor condition, while previous work often used metaphor-enhanced texts that lead to double explanations of a phenomenon. Furthermore, our actual comprehension question focused specifically on the quantum phenomenon, whereas other studies also tested comprehension of parts unrelated to the metaphor (see e.g., Alexander & Kulikowich, 1991; Jaeger & Wiley, 2015). Finally, different ways to measure actual comprehension exist, for instance through closed multiple-choice questions (Alexander & Kulikowich, 1991), open recall questions (Glynn & Takashi, 1998), inference questions (questions about 'what happens if something in the system changes'; Yanowitz, 2001) and prompting to write a full essay (Braasch & and Goldman, 2010; Jaeger & Wiley, 2015). These different ways probably also lead to different comprehension scores, with some tapping into more shallow levels of comprehension and others into deeper ones (see also Bernholt et al., 2023).

A closer look at the actual comprehension answers furthermore provides an interesting insight about the terms 'superposition' and 'entanglement'. Some participants wrote that superposition is about a competitive (market) position ($n$ = 46, 8.0%) and entanglement about intertwining things, such as weaving, mixing or melting ($n$ = 39, 6.6%). While the technical terms 'superposition' and 'entanglement' might not be considered metaphorical to quantum scientists (Beger & Smith, 2020), they may be processed differently by people outside the quantum field. Furthermore, only 13.1% of the participants in the metaphor condition reused the provided metaphors (a coin spinning in the air: $n$ = 33, 16.6%; rolling a pair of dice: $n$ = 19, 9.6%). Perhaps the metaphorical explanation remained too abstract or was too brief to fully understand the quantum phenomenon.

### 5.2 Effects of explanation type on attitude

We found no differences between the three conditions for affect-based attitudes  and cognition-based attitudes. This differs from the small persuasive effect of metaphors compared to literal messages as found in meta-analyses (O'Keefe & Hoeken, 2021; Sopory & Dillard, 2002; Van Stee, 2018). The target domain was likely too unfamiliar with low quantum technology awareness scores (see Figure A.1), making the cognitive effort required to process the metaphor too great (Sopory & Dillard, 2002; Van Stee, 2018).

Previous research suggests that using metaphors to explain quantum phenomena, such as superposition and entanglement, might improve attitudes towards quantum technology (Grinbaum, 2017). However, in our study, attitudes towards quantum technology were positive across conditions, regardless of whether a quantum phenomenon was explained. This suggests that resistance, found with other emergent technologies (Druckman & Bolsen, 2011; Kurath & Gisler, 2009), may currently not apply for quantum technology. Given the low awareness of quantum technology, it is plausible these positive attitudes formed during the survey (see also Druckman & Bolsen, 2011; Van Giesen et al., 2018). Apparently, reading a neutral-to-slightly-positive article about quantum technology is enough to form these attitudes, regardless of an additional explanation about underlying phenomena.

### 5.3 Comprehension as a mediator between explanation type and attitudes

Results of the mediation analyses suggest that explanations have a very small negative indirect effect on people's attitudes towards quantum technology, due to the fact that people feel they understood the news article less. This is consistent with Akin et al. (2021), who found that people's beliefs about their own knowledge of a technology were more influential in shaping their attitudes than their actual knowledge. While we also found a very small positive indirect effect of explanations on attitudes mediated by actual comprehension, this was counteracted by a direct negative effect. Apparently, improving people's actual comprehension of the underlying quantum phenomena does not necessarily result in more positive views towards quantum technology, as there could be other mechanisms at play that counteract the positive effect on comprehension.

In addition, and in contrast with what Grinbaum (2017) advocated, the type of explanation, metaphorical or non-metaphorical, did not lead to different outcomes on comprehension and therefore on attitudes. One possible explanation for this is that the metaphors used in our study, although deemed most accurate and valid, did not evoke a sufficiently vivid image in people's minds (Wyer Jr. & Shrum, 2015). It could well be that with a different metaphor attitudes would have been affected more, for example with the famous metaphor of Schrödinger's cat. While experts in preparation of the experiment acknowledged that the metaphor is very well-known (see also Van de Merbel et al., 2024) and therefore may resonate better, some experts pointed out that the metaphor is incorrect and too complicated to bring across.

*5.4 Resistance to metaphor*

Finally, we want to point out an interesting additional finding: some experts in preparation of the experiment indicated that the metaphors made the quantum phenomena unnecessarily complicated and mysterious, creating only more confusion. One expert even emailed us afterwards saying that their *"strong feeling is that it is dangerous to use a metaphor for such a profound phenomenon"*. This aligns with the body of research that metaphors can cause resistance. Such a comment also ties in with Feynman's quote that quantum phenomena are too complex to explain (Feynman, 1967), which could be bad for the democratization of quantum technology (Seskir et al., 2023).

*5.5 Practical implications*

In this study, we have shown that explanations of quantum phenomena in a news article affect people's comprehension, which subsequently mediates attitudes towards quantum technology. We recommend that, if the goal of a news article is to increase people's feeling that they have understood the text, journalists better skip the explanation of counterintuitive quantum phenomena. De Jong (2025) has taken this a step further by arguing that quantum phenomena should not be explained to a general public, but there should be an increased focus on the functional capacities of quantum technology to lower the barrier to engage in ethical discussions about quantum.

Our recommendation changes if the goal is to increase people's actual understanding of the underlying quantum phenomena of quantum technology. Explanations help to increase actual comprehension of quantum phenomena, but our findings suggest that there is no additional benefit of using metaphors to explain counterintuitive quantum phenomena.

**Notes**

1. Schrödinger's cat is a thought experiment devised by Erwin Schrödinger (Die gegenwärtige Situation in der Quantenmechanik, E. Schrödinger, 29 Nov, 1935) to illustrate the paradoxes of quantum mechanics under the Copenhagen interpretation, but it is now more commonly used as a metaphor to explain quantum mechanics in popular culture (https://en.wikipedia.org/wiki/Schr%C3%B6dinger%27s_cat).

2. Here we slightly deviated from the preregistration, given that we made the call to examine perceived comprehension of the text as a whole, informed by previous studies that measured perceived comprehension at the level of the entire text.

3. Note that for gender, the 'other' category was too small to meet the assumptions of the chi-square test of independence, which we therefore excluded.

# Acknowledgements

We want to thank the interdisciplinary community of quantum scientists, communication experts and linguists who joined the research symposium *The Role of Language in Engaging Audiences in New (Quantum) Technology* and gave feedback on the research plan that led to this manuscript. This work was funded by the Dutch Research Council (NWO) through a Spinoza grant awarded to R. Hanson (project number SPI 63-264), and supported by the Dutch National Growth Fund (NGF) as part of the Quantum Delta NL programme. Aletta Meinsma received a Research Visit grant from the Network Institute of the Vrije Universiteit Amsterdam to develop the study within the Language & Communication research group.

# Appendix

*A.1 Text of stimulus material and different conditions*

Participants were assigned to read a newspaper article about quantum containing either a metaphorical or non-metaphorical explanation of superposition or entanglement or no explanation at all. The original Dutch version (A.1.1) and the English version (A.1.2), translated with the help of Google Translate and checked by all authors, are found below.

*A.1.1 Original Dutch version*

**Nieuwe quantumcomputer voor Nederland**

Nederland krijgt een van de acht nieuwe quantumcomputers die de Europese Commissie laat bouwen. Hiermee hopen ze Europa een betere concurrentiepositie te geven op het gebied van quantumtechnologie.

*23 Oktober 2024*

De Nederlandse quantumcomputer kost 20 miljoen euro. Hij komt op het Amsterdam Science Park en wordt naar verwachting in de zomer van 2026 gebouwd. Quantumcomputers werken anders dan de computers die we nu gebruiken. Ze maken gebruik van de principes van de quantumfysica, een domein binnen de natuurkunde dat zich bezighoudt met de allerkleinste deeltjes. Een belangrijk kenmerk daarin is *superpositie/verstrengeling*. Quantumcomputers maken onder andere gebruik van *superpositie/verstrengeling*, waardoor ze in de toekomst mogelijk bepaalde problemen sneller kunnen oplossen dan onze huidige computers.

| | Item: Superpositie | Item: Verstrengeling |
|---|---|---|
| *Conditie:* *Control* | No text. (0 words) | No text. (0 words) |
| *Conditie:* *Niet-metaforisch* | Superpositie is een quantumfenomeen. Een deeltje in superpositie bevindt zich niet slechts in één toestand. Zolang het deeltje in superpositie is, is het in een combinatie van verschillende toestanden tegelijkertijd. Deze situatie blijft bestaan totdat we het deeltje meten. Pas als we een meting aan het deeltje doen komt het deeltje in één toestand. Deeltjes in de quantumwereld kunnen dus tegelijkertijd in meerdere toestanden bestaan totdat ze worden gemeten of waargenomen. Zo werkt het in de quantumwereld. (77 words) | Verstrengeling is een quantumfenomeen. Als twee deeltjes verstrengeld zijn, betekent dit dat hun toestanden op een bepaalde manier met elkaar verbonden zijn. Als je de toestand van het ene deeltje meet, weet je direct wat de toestand van het andere deeltje is. Zelfs als die deeltjes zich aan weerszijden van het universum bevinden. De toestand van het ene deeltje zorgt er dus voor dat de toestand van het deeltje waarmee het verstrengeld is vastligt. Zo werkt het in de quantumwereld. (80 words) |
| *Conditie:* *Metaforisch* | Superpositie is een quantumfenomeen. Een deeltje in superpositie kun je vergelijken met een muntje dat in de | Verstrengeling is een quantumfenomeen. Verstrengelde deeltjes kun je vergelijken met een paar |

lucht draait. Zolang het muntje draait lijkt het alsof het tegelijkertijd kop en munt is. Deze situatie blijft bestaan totdat we de munt op tafel slaan. Pas als we de munt op tafel slaan komt het muntje op kop óf munt terecht. Het muntje lijkt dus tegelijkertijd kop én munt totdat het wordt gemeten of waargenomen. Zo werkt het in de quantumwereld. (79 words) dobbelstenen waarbij de uitkomsten altijd hetzelfde zijn. Wanneer de ene dobbelsteen wordt gegooid, is de uitkomst van de andere dobbelsteen direct bepaald. Zelfs als deze aan de andere kant van de speeltafel wordt gegooid. De uitkomst van de ene dobbelsteen zorgt er dus voor dat de uitkomst van de dobbelsteen waarmee die verbonden is vastligt. Zo werkt het in de quantumwereld. (73 words)

De komst van een quantumcomputer is een belangrijk moment voor de concurrentiepositie van Nederland in quantumtechnologie. Maar wat de technologie precies voor jou en mij gaat betekenen is nog onduidelijk. De betrokken onderzoekers hopen hier met dit project meer over te weten te komen.

*A.1.2 Translated English version*

**New quantum computer for the Netherlands**

The Netherlands will receive one of eight new quantum computers that the European Commission is having built. They hope that this will give Europe a better competitive position in the field of quantum technology.

*23 October 2024*

The Dutch quantum computer costs 20 million euros. It will be located at the Amsterdam Science Park and is expected to be built in the summer of 2026. Quantum computers work differently than the computers we use now. They use the principles of quantum physics, a domain within physics that deals with the smallest particles. An important characteristic of this is *superposition/entanglement*. Quantum computers use *superposition/entanglement*, among other things, as a result of which, in the future, they may be able to solve certain problems faster than our current computers can.

|  | *Item: Superposition* | *Item: Entanglement* |
|---|---|---|
| *Condition: Control* | No text. | No text. |
| *Condition: Non-metaphorical* | Superposition is a quantum phenomenon. A particle in superposition is not just in one state. As long as the particle is in superposition, it is in a combination of different states at the same time. This situation continues until we measure the particle. Only when we perform a measurement on the particle does the | Entanglement is a quantum phenomenon. When two particles are entangled, it means that their states are somehow connected. If you measure the state of one particle, you immediately know what the state of the other particle is. Even if those particles are on opposite sides of the universe. So the state of one particle |
| | particle enter a single state. Particles in the quantum world can therefore exist in multiple states at the same time until they are measured or observed. This is how the quantum world works. | ensures that the state of the particle it is entangled with is fixed. This is how the quantum world works. |
| *Condition: Metaphorical* | Superposition is a quantum phenomenon. A particle in superposition can be compared to a coin spinning in the air. As long as the coin is spinning, it appears to be heads and tails at the same time. This situation continues until we slap the coin on the table. Only when we slap the coin on the table does the coin land on either heads or tails. The coin therefore appears to be heads and tails at the same time until it is measured or observed. That is how it works in the quantum world. | Entanglement is a quantum phenomenon. Entangled particles can be compared to a pair of dice where the outcomes are always the same. When one die is thrown, the outcome of the other die is immediately determined. Even if it is thrown on the other side of the gaming table. The outcome of one die therefore ensures that the outcome of the die it is connected to is fixed. That is how it works in the quantum world. |

The arrival of a quantum computer is an important moment for the competitive position of the Netherlands in quantum technology. But what the technology will mean exactly for you and me is still unclear. The researchers involved hope to find out more about this with this project.

*A.2 Intercoder reliability on actual comprehension*

To ensure the reliability of the results on actual comprehension, two coders discussed the first 50 answers and rated the answers together. The two coders agreed in all cases and afterwards independently coded a sample of 10% of the answers (n = 117) to calculate intercoder agreement. We found perfect to near-perfect agreement between the coders for superposition and entanglement with alpha values between 0.86 and 1 (percent agreements between 96% - 100%), except for 2) there is a connection/correlation (alpha = 0.54, 84%). We modified this code slightly by specifying that using a synonym of connection/correlation, such as cooperation or fusion, would also be awarded a point. Furthermore, no point would be awarded if the answer only mentioned connection/correlation, without specifying that it was between something, such as small particles. Afterwards, the first coder coded the remaining 1,009 answers.

*A.3 Control variables*

**Awareness of quantum.** Prior awareness of quantum technology can influence comprehension when participants use it as a cue for assessment (Thiede et al., 2010) and can furthermore influence participants' attitudes with more aware participants holding more favourable attitudes (see Scheufele & Lewenstein, 2005 in the case of nanotechnology). To control for this effect, participants were asked (Scheufele & Lewenstein, 2005): *"How much have you heard, read or seen about the following technologies?" "Solar energy, Biotechnology, Quantum technology, Artificial Intelligence, Vaccines"* [1 = nothing at all, 7 = very much]. To mask the fact that we were only interested in awareness of quantum technology, we used 4 extra items shown in a random order, based on new technologies that were asked in the Special Eurobarometer 516 (European Commission, Directorate-General for Communication, 2021). In line with our expectations, awareness of quantum technology in the sample was low ($M$ = 2.78, $SD$ = 1.60), with 66.1% of participants ($n$ = 773) indicating they had not heard about quantum technology at all or little [scores: 1-3]. Compared to the other new technologies that we asked about, participants scored on average lowest for awareness of quantum technology (see Figure A.1).

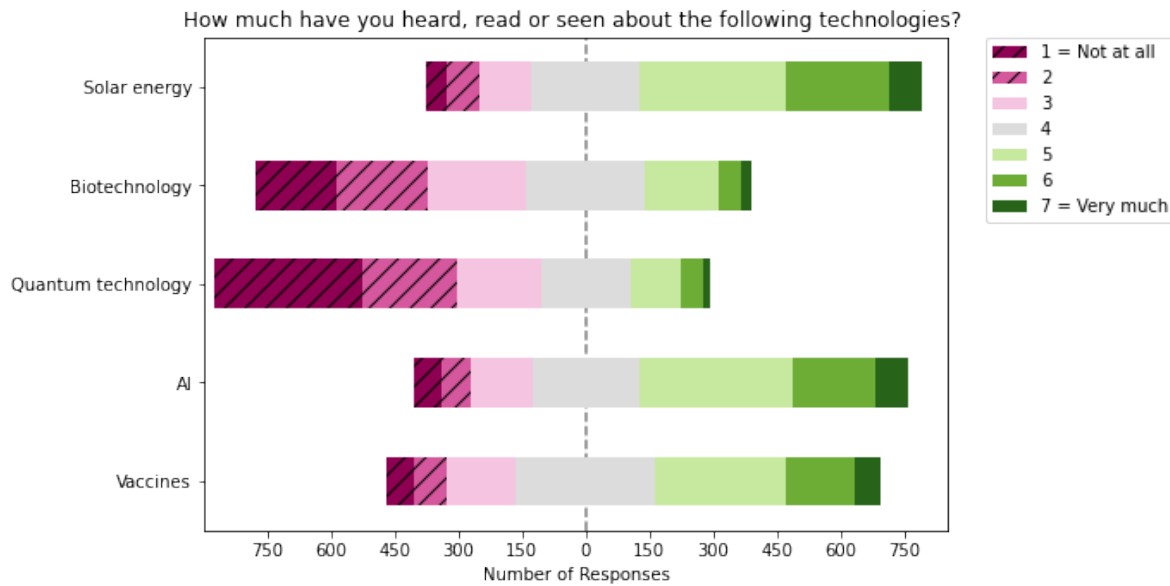

**Figure A.1.** Awareness of quantum technology in comparison to other new technologies. The plots are centred around the neutral (Likert score of 4) value. (Figure created through adaptation of the Github file: https://github.com/erikvansebille/KH_personalization_effect/tree/main).

**Science news use.** People who tend to read the science news section in newspapers (the most popular section in which articles on quantum science and technology appear; Meinsma et al., 2025) might hold more positive views on quantum technology compared to people who skip this section (see e.g., Scherrer, 2023; Scheufele & Lewenstein, 2005). Therefore, we measured participants' science news use with the item (Eurobarometer, 2023): *"Thinking about news and other information, which of the following topics have you accessed in the past 7 days? (Multiple answers allowed)": "Local news, National politics, European and international affairs, Sport, Crime and accidents, Financial and economic news, Science and technology, Hobbies and lifestyle, Culture and art, People (e.g., royals, celebrities, etc.), Other, None, Don't know".* We found a total of $n$ = 344 (29.5%) had accessed science news in the past 7 days, which is slightly higher than the 24% found in 2023 by the Eurobarometer (Eurobarometer, 2023).

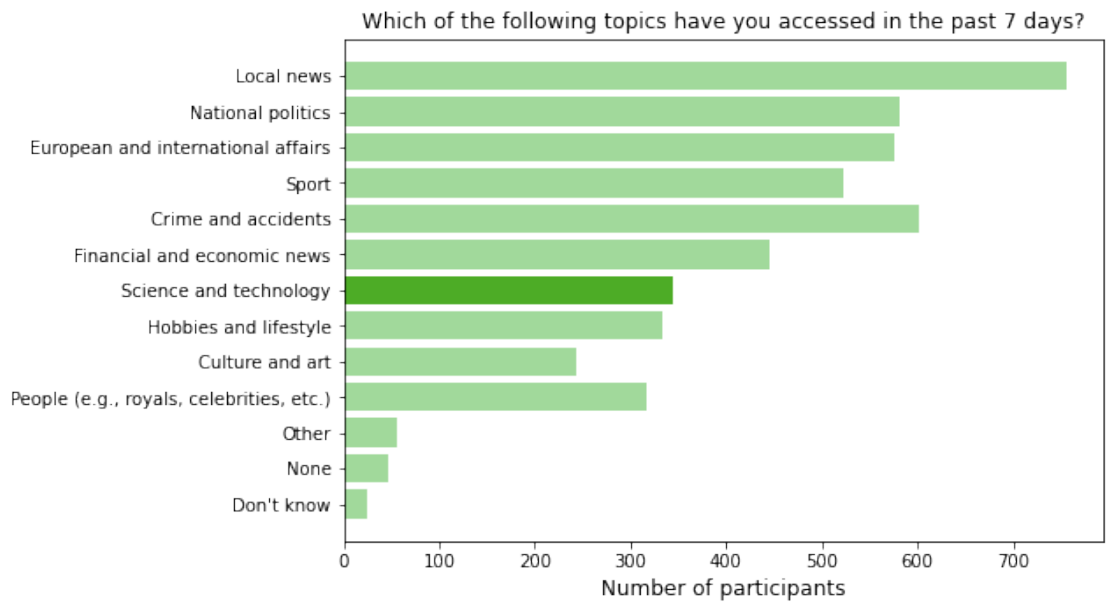

**Figure A.2.** News use of the participants in our sample, where science news use indicates the answer 'Science and technology'.

**Interest in new technology.** People may use interest in a topic as a cue to judge their comprehension of a text (Thiede et al., 2010). We measured participants' interest in new technologies with 6 items adapted from Shulman et al., 2020 ("science technologies" was substituted for "new technologies"): *"I am interested in learning about new technologies", "I find the debate surrounding new technologies interesting", "I want to learn more about new technologies", "New technologies are exciting", "I find new technology boring"* (reverse coded), *"New technological ideas are thought-provoking"* [1 = strongly disagree, 7 = strongly agree]. These six items were averaged into an index (*M* = 4.40, *SD* = 1.29, Cronbach's alpha = 0.91).

**Faith in intuition.** People who have a high faith in intuition tend to rely more on affect (van Giesen et al., 2018). Therefore, we measured participants' faith in intuition with 5 items (Epstein et al., 1996): *"I trust my initial feelings about people", "I believe in trusting my hunches", "My initial impressions of people are almost always right", "When it comes to trusting people, I can usually rely on my "gut feelings." ", "I can usually feel when a person is right or wrong even if I can't explain how I know."* [1 = completely false, 7 = completely true]. The five items were averaged into an index (*M* = 4.87, *SD* = 0.95, Cronbach's alpha = 0.86).

**Need for cognition.** People who have a high need for cognition rely more on cognition (van Giesen et al., 2018). The 5 items were (Epstein et al., 1996): *"I don't like to have to do a lot of thinking"* (reverse coded), *"I try to avoid situations that require thinking in depth about something"* (reverse coded), *"I prefer to do something that challenges my thinking abilities rather than something that requires little thought", "I prefer complex to simple problems", "Thinking hard and for a long time about something gives me little satisfaction* (reverse coded)"* [1 = completely false, 7 = completely true]. The five items were averaged into an index (*M* = 4.57, *SD* = 1.05, Cronbach's alpha = 0.75).

*A.4 Tables for the mediation analyses*

**Table A.1.**

The indirect, component (path a and b), direct (path c') and total effects from the mediation analyses with perceived comprehension (PC) as a mediator. The contrasts used are: 1 vs. 2 = control -- non-metaphorical and 1 vs. 3 = control -- metaphorical. 'Cogn.' refers to cognition.

| | | | | Non-metaphorical – Control, PC as mediator | | | | |
| | | | | 95% C.I. | | | | |
| Type | Effect | Estimate | SE | Lower | Upper | β | z | p |
|---|---|---|---|---|---|---|---|---|
| Indirect | 1 vs. 2 ⇒ PC ⇒ Affect | -0.13 | 0.04 | -0.21 | -0.06 | -0.07 | -3.64 | < .001 |
| | 1 vs. 2 ⇒ PC ⇒ Cogn. | -0.12 | 0.03 | -0.19 | -0.05 | -0.06 | -3.52 | < .001 |
| Path a | 1 vs. 2 ⇒ PC | -0.32 | 0.09 | -0.48 | -0.15 | -0.12 | -3.73 | < .001 |
| Path b | PC ⇒ Affect | 0.42 | 0.02 | 0.39 | 0.46 | 0.55 | 21.33 | < .001 |
| | PC ⇒ Cogn. | 0.38 | 0.02 | 0.33 | 0.42 | 0.46 | 16.48 | < .001 |
| Path c' | 1 vs. 2 ⇒ Affect | 0.09 | 0.05 | -0.01 | 0.21 | 0.05 | 1.76 | 0.079 |
| | 1 vs. 2 ⇒ Cogn. | 0.04 | 0.06 | -0.09 | 0.17 | 0.02 | 0.65 | 0.519 |
| Total | 1 vs. 2 ⇒ Affect | -0.04 | 0.07 | -0.17 | 0.09 | -0.02 | -0.58 | 0.565 |
| | 1 vs. 2 ⇒ Cogn. | -0.08 | 0.07 | -0.22 | 0.06 | -0.04 | -1.11 | 0.266 |

| | | | | Metaphorical – Control, PC as mediator | | | | |
| | | | | 95% C.I. | | | | |
| Type | Effect | Estimate | SE | Lower | Upper | β | Z | p |
|---|---|---|---|---|---|---|---|---|
| Indirect | 1 vs. 3 ⇒ PC ⇒ Affect | -0.12 | 0.04 | -0.20 | -0.05 | -0.06 | -3.37 | < .001 |
| | 1 vs. 3 ⇒ PC ⇒ Cogn. | -0.11 | 0.03 | -0.18 | -0.05 | -0.05 | -3.28 | .001 |
| Path a | 1 vs. 3 ⇒ PC | -0.29 | 0.08 | -0.46 | -0.13 | -0.11 | -3.41 | < .001 |
| Path b | PC ⇒ Affect | 0.42 | 0.02 | 0.39 | 0.46 | 0.55 | 21.33 | < .001 |
| | PC ⇒ Cogn. | 0.38 | 0.02 | 0.33 | 0.42 | 0.46 | 16.48 | < .001 |
| Path c' | 1 vs. 3 ⇒ Affect | 0.01 | 0.05 | -0.10 | 0.12 | 0.01 | 0.22 | 0.826 |
| | 1 vs. 3 ⇒ Cogn. | -0.02 | 0.06 | -0.14 | 0.10 | -0.01 | -0.38 | 0.708 |
| Total | 1 vs. 3 ⇒ Affect | -0.11 | 0.07 | -0.24 | 0.01 | -0.06 | -1.7 | 0.089 |
| | 1 vs. 3 ⇒ Cogn. | -0.13 | 0.07 | -0.26 | 0.00 | -0.06 | -1.96 | 0.050 |

*Note.* Indirect = Path a * Path b, Total = Indirect + Path c', z-statistic = Estimate / SE.
*Note.* Betas are completely standardized effect sizes.

**Table A.2.**

The indirect, component (path a and b), direct (path c') and total effects from the mediation analyses with actual comprehension (AC) as a mediator. The contrasts used are: 1 vs. 2 = control -- non-metaphorical and 1 vs. 3 = control -- metaphorical. 'Cogn.' refers to cognition.

| | | Non-metaphorical – Control, AC as mediator | | | | | | |
|---|---|---|---|---|---|---|---|---|
| | | | | 95% C.I. | | | | |
| Type | Effect | Estimate | SE | Lower | Upper | β | z | p |
| Indirect | 1 vs. 2 ⇒ AC ⇒ Affect | 0.15 | 0.02 | 0.11 | 0.19 | 0.07 | 6.91 | < .001 |
| | 1 vs. 2 ⇒ AC ⇒ Cogn. | 0.15 | 0.02 | 0.11 | 0.20 | 0.07 | 6.39 | < .001 |
| Path a | 1 vs. 2 ⇒ AC | 0.43 | 0.05 | 0.33 | 0.54 | 0.26 | 8.22 | < .001 |
| Path b | AC ⇒ Affect | 0.34 | 0.03 | 0.27 | 0.40 | 0.28 | 10.26 | < .001 |
| | AC ⇒ Cogn. | 0.35 | 0.04 | 0.27 | 0.42 | 0.27 | 9.40 | < .001 |
| Path c' | 1 vs. 2 ⇒ Affect | -0.18 | 0.07 | -0.31 | -0.05 | -0.09 | -2.71 | 0.007 |
| | 1 vs. 2 ⇒ Cogn. | -0.23 | 0.07 | -0.37 | -0.09 | -0.11 | -3.32 | < .001 |
| Total | 1 vs. 2 ⇒ Affect | -0.04 | 0.07 | -0.17 | 0.09 | -0.02 | -0.57 | 0.566 |
| | 1 vs. 2 ⇒ Cogn. | -0.08 | 0.07 | -0.22 | 0.06 | -0.04 | -1.11 | 0.266 |
| | | Metaphorical – Control, AC as mediator | | | | | | |
| | | | | 95% C.I. | | | | |
| Type | Effect | Estimate | SE | Lower | Upper | β | z | p |
| Indirect | 1 vs. 3 ⇒ AC ⇒ Affect | 0.10 | 0.02 | 0.07 | 0.14 | 0.05 | 5.77 | < .001 |
| | 1 vs. 3 ⇒ AC ⇒ Cogn. | 0.11 | 0.02 | 0.07 | 0.15 | 0.05 | 5.65 | < .001 |
| Path a | 1 vs. 3 ⇒ AC | 0.31 | 0.05 | 0.21 | 0.40 | 0.19 | 6.32 | < .001 |
| Path b | AC ⇒ Affect | 0.34 | 0.03 | 0.27 | 0.40 | 0.28 | 10.26 | < .001 |
| | AC ⇒ Cogn. | 0.35 | 0.04 | 0.27 | 0.42 | 0.27 | 9.40 | < .001 |
| Path c' | 1 vs. 3 ⇒ Affect | -0.22 | 0.07 | -0.35 | -0.09 | -0.11 | -3.31 | < .001 |
| | 1 vs. 3 ⇒ Cogn. | -0.24 | 0.07 | -0.37 | -0.11 | -0.11 | -3.55 | < .001 |
| Total | 1 vs. 3 ⇒ Affect | -0.11 | 0.07 | -0.24 | 0.02 | -0.06 | -1.66 | 0.097 |
| | 1 vs. 3 ⇒ Cogn. | -0.13 | 0.07 | -0.27 | 0.00 | -0.06 | -1.94 | 0.053 |

*Note.* Indirect = Path a * Path b, Total = Indirect + Path c', z-statistic = Estimate / SE
*Note.* Betas are completely standardized effect size

*A.4 Expert insights study*

Table A.3 shows the metaphors presented to the experts in the expert insights study in preparation of the experiment. Note that we have slightly modified the wording of the coin and dice metaphor in the experiment in a few places compared to the text presented to the experts, in order to make the texts as similar as possible in structure to the non-metaphorical version of the text. However, the metaphorical content has not changed.

**Table A.3.** The metaphors used in the expert insights study, which were generated by ChatGPT 3.5 and modified to make sure the structure of each metaphor is similar.

| | | | Superposition |
|---|---|---|---|
| | Metaphor | Original Dutch text | Translated English text |
| 1 | A coin spinning in the air | Quantumsuperpositie is als een munt die in de lucht draait. Totdat we de munt op tafel slaan, is het alsof de munt tegelijkertijd kop en munt is. Echter, zodra we de munt op tafel slaan, komt de munt op kop óf munt terecht. Dit illustreert hoe deeltjes in de quantumwereld tegelijkertijd in meerdere toestanden kunnen bestaan totdat ze worden gemeten of waargenomen, net zoals een munt zowel kop als munt is totdat we de munt op tafel slaan. | Quantum superposition is like a coin spinning in the air. Until we hit the coin on the table, it is as if the coin is both heads and tails at the same time. However, as soon as we hit the coin on the table, the coin lands either heads or tails. This illustrates how particles in the quantum world can exist in multiple states at the same time until they are measured or observed, just as a coin is both heads and tails until we hit it on the table. |
| 2 | A radio producing a jumble of sounds | Quantumsuperpositie is als een radio die een wirwar aan geluiden voortbrengt. Totdat de radio-ontvanger aan een zender is gekoppeld, is het alsof de radio tegelijkertijd op alle mogelijke zenders is afgestemd. Echter, zodra de radio-ontvanger aan een zender is gekoppeld, stemt de radio meteen nog maar op één van de mogelijke zenders af. Dit illustreert hoe deeltjes in de quantumwereld tegelijkertijd in meerdere toestanden kunnen bestaan totdat ze worden gemeten of waargenomen, net zoals de radio op alle mogelijke zenders is afgestemd totdat de radio-ontvanger aan een zender is gekoppeld. | Quantum superposition is like a radio that produces a jumble of sounds. Until the radio receiver is hooked up to a transmitter, it is as if the radio is tuned to all possible channels at once. However, as soon as the radio receiver is hooked up to a transmitter, the radio immediately tunes to only one of the possible channels. This illustrates how particles in the quantum world can exist in multiple states at once until they are measured or observed, just as the radio is tuned to all possible channels until the radio receiver is hooked up to a transmitter. |
| 3 | A cat, a vial of poison | Quantumsuperpositie is als een kat, een flesje gif en een radioactief atoom in een afgesloten doos. Totdat we de doos | Quantum superposition is like a cat, a vial of poison, and a radioactive atom in a sealed box. Until we open the box, it is as if |
| | and a radioactive atom in a locked box | openen, is het alsof de kat tegelijkertijd levend en dood is. Echter, zodra we de doos openen zien we dat óf het atoom is vervallen waardoor het flesje gif is gebroken en de kat is gedood, óf dat het atoom niet is vervallen en de kat levend is. Dit illustreert hoe deeltjes in de quantumwereld tegelijkertijd in meerdere toestanden kunnen bestaan totdat ze worden gemeten of waargenomen, net zoals een kat zowel levend als dood is totdat we de doos openen. | the cat is both alive and dead at the same time. However, once we open the box, we see that either the atom has decayed, breaking the vial of poison and killing the cat, or the atom has not decayed and the cat is alive. This illustrates how particles in the quantum world can exist in multiple states at the same time until they are measured or observed, just as a cat is both alive and dead until we open the box. |
| 4 | An artist dabbing his brush in multiple colours | Quantumsuperpositie is als een kunstenaar die zijn penseel in meerdere kleuren dept. Totdat het penseel het canvas raakt, is het alsof het penseel tegelijkertijd in alle mogelijke kleurencombinaties aanwezig is. Echter, zodra het penseel het canvas raakt, laat het canvas meteen nog maar één van de mogelijke kleurencombinaties zien. Dit illustreert hoe deeltjes in de quantumwereld tegelijkertijd in meerdere toestanden kunnen bestaan totdat ze worden gemeten of waargenomen, net zoals het penseel zich in alle mogelijke kleurencombinaties bevindt totdat het penseel het canvas raakt. | Quantum superposition is like an artist dabbing his brush in multiple colors. Until the brush touches the canvas, it is as if the brush is in all possible color combinations at the same time. However, as soon as the brush touches the canvas, the canvas immediately shows only one of the possible color combinations. This illustrates how particles in the quantum world can exist in multiple states at the same time until they are measured or observed, just as the brush is in all possible color combinations until it touches the canvas. |
| 5 | A musician composing a music piece | Quantumsuperpositie is als een muzikant die een muziekstuk componeert en meerdere muzieknoten op hetzelfde blad plaatst. Totdat de muzikant de partituur speelt, is het alsof de muziek zich tegelijkertijd in alle mogelijke melodieën bevindt. Echter, zodra de muzikant de partituur speelt, bevindt de muziek zich meteen nog maar in één van de mogelijke melodieën. Dit illustreert hoe deeltjes in de quantumwereld tegelijkertijd in meerdere toestanden kunnen bestaan totdat ze worden gemeten of waargenomen, net zoals de muziek zich in alle mogelijke melodieën bevindt totdat de muzikant de partituur speelt. | Quantum superposition is like a musician composing a piece of music and placing multiple notes on the same sheet of paper. Until the musician plays the score, it is as if the music is in all possible melodies at once. However, as soon as the musician plays the score, the music is immediately in only one of the possible melodies. This illustrates how particles in the quantum world can exist in multiple states at once until they are measured or observed, just as the music is in all possible melodies until the musician plays the score. |

| | | | Entanglement |
|---|---|---|---|
| | Metaphor | Original Dutch text | Translated English text |
| 1 | A pair of dice | Quantumverstrengeling is als het gooien van een paar dobbelstenen. Wanneer de ene dobbelsteen wordt gegooid, is de uitkomst van de andere dobbelsteen vooraf bepaald, zelfs als deze aan de andere kant van de speeltafel wordt gegooid. Dit illustreert hoe in de quantumwereld de toestand van het ene deeltje de toestand bepaalt van het deeltje waarmee het verstrengeld is, net zoals de uitkomst van de ene dobbelsteen de uitkomst van de andere dobbelsteen bepaalt. | Quantum entanglement is like rolling a pair of dice. When one die is rolled, the outcome of the other die is predetermined, even if it is rolled on the other side of the gaming table. This illustrates how in the quantum world, the state of one particle determines the state of the particle it is entangled with, just as the outcome of one die determines the outcome of the other die. |
| 2 | Two dancers performing a perfectly synchronized dance routine | Quantumverstrengeling is als twee dansers die een perfect gesynchroniseerde dansroutine uitvoeren. Wanneer de ene danser beweegt, beweegt de andere danser op een gecoördineerde manier alsof ze een onzichtbare link delen, zelfs als ze zich aan andere kanten van het podium bevinden. Dit illustreert hoe in de quantumwereld de toestand van het ene deeltje de toestand bepaalt van het deeltje waarmee het verstrengeld is, net zoals de bewegingen van de ene danser de bewegingen van de andere danser bepalen. | Quantum entanglement is like two dancers performing a perfectly synchronized dance routine. When one dancer moves, the other dancer moves in a coordinated manner as if they share an invisible link, even if they are on opposite sides of the stage. This illustrates how in the quantum world, the state of one particle determines the state of the particle it is entangled with, just as the movements of one dancer determine the movements of the other dancer. |
| 3 | A telepathic twin | Quantumverstrengeling is als een telepathische tweeling die onmiddellijk elkaars gedachten kan kennen. Wanneer de ene tweeling van gedachten verandert, verandert de ander onmiddellijk ook van gedachten, zelfs als ze zich aan andere kanten van de planeet bevinden. Dit illustreert hoe in de quantumwereld de toestand van het ene deeltje de toestand bepaalt van het deeltje waarmee het verstrengeld is, net zoals de gedachten van de ene tweeling de gedachten van de andere tweeling bepalen. | Quantum entanglement is like a pair of telepathic twins who can instantly know each other's thoughts. When one twin changes their mind, the other instantly changes their mind too, even if they are on opposite sides of the planet. This illustrates how in the quantum world, the state of one particle determines the state of the particle it is entangled with, just as the thoughts of one twin determine the thoughts of the other twin. |
| 4 | Two compass | Quantumverstrengeling is als het hebben van twee kompasnaalden die altijd in tegengestelde richtingen wijzen. | Quantum entanglement is like having two compass needles that always point in opposite directions. When you turn one needle |
| | needles that always point in opposite directions | Wanneer je de ene naald naar het noorden draait, wijst de andere onmiddellijk naar het zuiden, ook al bevinden ze zich aan de andere kant van de zeilboot. Dit illustreert hoe in de quantumwereld de toestand van het ene deeltje de toestand bepaalt van het deeltje waarmee het verstrengeld is, net zoals het draaien van de naald van het ene kompas de naald van het andere kompas aanpast. | north, the other immediately points south, even though they are on opposite sides of the sailboat. This illustrates how in the quantum world, the state of one particle determines the state of the particle it is entangled with, just as turning the needle of one compass adjusts the needle of the other compass. |
| 5 | Two clocks with perfectly synchronized second hands | Quantumverstrengeling is als twee klokken waarvan de secondewijzers perfect gesynchroniseerd zijn. Wanneer u de tijd op de ene klok wijzigt, past de andere zich onmiddellijk aan, zelfs als deze zich aan de andere kant van de kamer bevindt. Dit illustreert hoe in de quantumwereld de toestand van het ene deeltje de toestand bepaalt van het deeltje waarmee het verstrengeld is, net zoals het veranderen van de tijd op de ene klok de tijd op de andere klok aanpast. | Quantum entanglement is like two clocks with perfectly synchronized second hands. When you change the time on one clock, the other clock immediately adjusts, even if it is across the room. This illustrates how in the quantum world, the state of one particle determines the state of the particle it is entangled with, just as changing the time on one clock changes the time on the other clock. |

*A.5 The variety of explanations from experts for their rankings*

Table A.4 shows the variety of answers given to the question: "If you were to use a comparison in a conversation with non-quantum experts about quantum superposition/quantum entanglement, which of these would you use? Rank them in order from most likely to least likely."

**Table A.4.** Answers to the question 'Please explain your ranking'. Answers are grouped, translated and edited for clarity.

| | | | Superposition |
|---|---|---|---|
| | Metaphor | Positive | Negative |
| 1 | A coin spinning in the air | 1. The metaphor best illustrates that the outcome of the measurement is indeterminate until a measurement is performed.<br>2. The metaphor provides the most minimalistic and simplest explanation.<br>3. The metaphor lends itself best to subsequently explain other concepts such as entanglement, measurement, quantum key distribution etc.<br>4. The metaphor best illustrates that a measurement provides a kind of probability of different outcomes.<br>5. The metaphor appeals to the imagination of people and appeals to the largest audience. | 1. The metaphor explains what chance is (classical statistics), not what superposition is.<br>2. The metaphor does not take into account that you can change bases. |
| 2 | A radio producing a jumble of sounds | 1. The metaphor best illustrates that superpositions are real, in the sense that they were not simply randomly generated.<br>2. The metaphor is most accurate, because it explicitly places the importance of a measurement in the description. | 1. The metaphor brings with it all sorts of nuances that only confuse people more.<br>2. The metaphor is not correct.<br>3. The metaphor is incomprehensible.<br>4. The metaphor is too specific, which means it only appeals to the imagination of a limited number of people. |
| 3 | A cat, a vial of poison and a radioactive atom in a locked box | 1. The metaphor is very well known and therefore resonates better.<br>2. The metaphor is nice to explain entanglement. | 1. The metaphor is not correct.<br>2. The metaphor is too complicated to convey.<br>3. The metaphor is too specific, which means it only appeals to the imagination of a limited number of people. |

| 4 | An artist dabbing his brush in multiple colours | | 1. The metaphor brings all kinds of nuances that only confuse people more.<br>2. The metaphor unnecessarily brings the problem of mixing colour into it.<br>3. The metaphor is not valid.<br>4. The metaphor does not illustrate the quantum aspect enough, because if you dab a brush in multiple colours, there will probably be multiple colours on the canvas.<br>5. The metaphor is too specific, which means that it only appeals to the imagination of a limited number of people. |
| 5 | A musician composing a music piece | 1. The metaphor is valid. | 1. The metaphor brings with it all kinds of nuances that only confuse people more.<br>2. The metaphor is incomprehensible / complicated because of the many outcomes.<br>3. The metaphor does not illustrate the quantum aspect enough, because the musician can choose what he is going to play (in a quantum measurement we cannot choose what the outcome of the measurement will be).<br>4. The metaphor is worded too unclearly.<br>5. The metaphor is too specific, which means that it only appeals to the imagination of a limited number of people. |

| | | | Entanglement |
|---|---|---|---|
| | Metaphor | Positive | Negative |
| 1 | A pair of dice | 1. The metaphor illustrates that the outcome is not known in advance. 2. The metaphor illustrates that a 'measurement' is performed on one part of the entangled state which determines the state of the other part. 3. The metaphor illustrates the probabilistic collapse of the wave function of an entangled state. 4. The metaphor illustrates the 'random' character. 5. The metaphor illustrates that probability is a description of our knowledge. 6. The metaphor is very simple and precise. | 1. The metaphor says that the state of the other die is known "automatically", while the states of the two dice are correlated in a way that is not classically possible. 2. The metaphor does not illustrate the superposition part, namely that both particles are in different states at the same time, until it is determined. 3. The metaphor misses the instantaneous. |
| 2 | Two dancers performing a perfectly synchronized dance routine | | 1. The metaphor implies communication by comparing it to humans, while one of the most common misconceptions about entanglement is that you can communicate faster than the speed of light. 2. The dancers can rehearse the dance in advance, so it is not strange that the dancers dance exactly in (anti)phase, that is routine. People can consciously perform actions and synchronize that. 3. The metaphor is too vague. 4. The metaphor suggests 'spooky action at a distance', which is the most misunderstood aspect of entanglement. |
| 3 | A telepathic twin | 1. The metaphor uses a much larger distance (other side of the planet). 2. The metaphor has somewhere the idea that there is not necessarily a predetermined local variable (at least, if we assume free will etc.). 3. The metaphor emphasizes the immediate nature of entanglement (the state of the second is determined immediately). | 1. The 'determination' in the mind is disturbing. 2. The metaphor implies communication by comparing it to people, while one of the most common misconceptions about entanglement is that you can communicate faster than the speed of light. 3. Suggesting telepathy is not ideal. |
| | | | 4. The metaphor suggests 'spooky action at a distance', which is the most misunderstood aspect of entanglement.<br>5. Two people can consciously perform actions and then synchronize them.<br>6. The element of chance is missing where a measurement in one place projects the entire quantum entangled system from an 'indeterminate state' to a combined fixed final state. |
| 4 | Two compass needles that always point in opposite directions | 1. The metaphor is intuitively clearest because the action does not come from within, but is commanded externally. | 1. The state is always visible and so there is a kind of analogue state.<br>2. The compasses could still be linked via a magnetic field.<br>3. The metaphor describes a kind of non-existent correlation that is not quantum mechanical, but also not classical.<br>4. The metaphor suggests 'spooky action at a distance', which is the most misunderstood aspect of entanglement.<br>5. Entanglement cannot be used to cause changes at a distance, like a compass needle that you turn north. |
| 5 | Two clocks with perfectly synchronized second hands | 1. The metaphor is nice.<br>2. The metaphor is intuitively the clearest, because the action does not come from within, but is externally ordered. | 1. This gives all sorts of issues with the theory of relativity.<br>2. The time in both clocks is a very clear local variable, and therefore resembles most 'normal' classical correlation. The state is always visible and therefore there is a kind of analogue state.<br>3. The metaphor describes a kind of non-existent correlation that is not quantum mechanical, but also not classical.<br>4. The superposition is missing.<br>5. If two particles are entangled, one of the particles can still be adjusted independently of the other by means of, in jargon, a "local transformation". What changes is the correlation. Hence, an active |

change on one particle influences the other particle does not apply.

6. The metaphor suggests 'spooky action at a distance', which is the most misunderstood aspect of entanglement.

7. Entanglement cannot be used to cause changes at a distance, like changing the time of a clock.